# Catalogues, parameters and distributions of orbital binaries

Oleg Malkov[1,2], Dmitry Chulkov[1]

[1] Institute of Astronomy of the Russ. Acad. Sci., 48 Pyatnitskaya Street, Moscow 119017, Russia
[2] Faculty of Physics, Moscow State University, Moscow 119992, Russia

Abstract: The most complete list of visual binary systems with known orbital elements is compiled. It is based on OARMAC and ORB6 data and contains 3139 orbits for 2278 pairs. A refined subset of high quality orbits with available distance was also compiled. Relations between and distributions along different observational parameters are constructed, and an analysis of selection effects is made. Dynamical, photometric and spectral masses of systems are estimated, and reasons for discrepancies between them are discussed.

1. Introduction

Orbital binary is a visual binary with known orbital elements (and known distance). Orbital binaries are essential objects for determining dynamical and physical properties of stars, especially masses, through a combined analysis of photometric and astrometric data. Along with double-lined eclipsing binaries (see details in Popper, 1980; Torres et al. 2010), orbital binaries with known distances are the only types of detached binary systems that enable one to determine stellar masses and semi-major axes of orbits.

Orbital binaries are quite numerous: current (May 2013) versions of principal catalogues of orbital binaries contain data on 2200 (Observatorio Astronomico Ramon Maria Aller Catalogue of Orbits and Ephemerides of Visual Double Stars, hereafter OARMAC, Docobo et al. 2001; Docobo et al. 2012a; Docobo et al. 2012b, http://www.usc.es/astro/catalog.htm) to 2400 orbits (Sixth Catalog of Orbits of Visual Binary Stars, hereafter ORB6, Hartkopf et al. 2001, http://ad.usno.navy.mil/wds/orb6.html). However, the former catalogue has an uncomfortable format, contains outdated photometry and lacks orbital parameter uncertainties. The latter catalogue contains neither spectral classification nor parallax information, and the period and semi-major axis are not provided for a few stars. Although they both represent the best collections of published data on orbital binaries, they are not exhaustive.

Other catalogues of orbital binaries contain many fewer objects and much less data. Abt (2005) provides eccentricity for 391 binaries (including spectroscopic). Orbital elements and masses for some 200 nearby visual binaries are given in Soederhjelm's (1999) catalogue.

In this work we present a compilation of orbital binaries, drawn from OARMAC and ORB6 and their statistical analysis. In Section 2 we briefly describe a process of the orbital binaries list compilation. In Section 3 we estimate dynamical masses of orbital binaries and, using magnitudes and spectral types of components, compare them with photometric and spectral masses. Reasons for discrepancies between masses derived by various methods are also discussed here. In Section 4 we present some relations between and distributions along observational parameters of orbital binaries and discuss selection effects. Finally, in Section 5 we draw conclusions.

2. The list of orbital binaries compilation and refinement

To compile the orbit list, we combined data from both OARMAC and ORB6. At this stage, we maintained systems without parallax, but removed systems without a period / semi-major axis. The resulting list contains 3139 orbits for 2278 pairs. Those 2278 pairs combine into 2016 binaries, 76 triples, 26 quadruples, 5 quintuples, and 2 septuples. Recent photometry and spectral classification, when absent, was added from WDS. The following data were added from SIMBAD: magnitudes and spectral types (if absent from the original catalogues and WDS), the parallax with its corresponding uncertainty (taken from van Leeuwen, 2007), and an indication of the spectroscopic and eclipsing nature of systems. In 650 records of the orbit list there is no available photometry for the secondary component, in 65 records there is no available spectral type, and the parallax for 270 records is unknown, zero, or negative. The procedure of compilation and the format of the resulting list are described in detail in (Malkov et al. 2012).

To carry out a statistical analysis of orbital binaries, one needs to "refine" the resulting orbit list. We removed triples and systems of higher multiplicity to avoid biases in dynamical mass distribution. We also removed systems with an unknown, zero, or negative parallax. Poor-quality orbits and astrometric orbits (ORB6 quality grades 4, 5 or 9 and OARMAC quality grade C) were also removed from the refined list. This refined list contains orbits of 652 systems. Spectral type of at least one component is available for all systems. In 35 records of the list there is no available photometry for the secondary component (see Appendix A).

3. Dynamical, photometric and spectral masses

The resulting orbit list often contains several orbits per pair, which is not convenient for statistical studies. Because the quality grades of these orbits are usually equal and both catalogues do not provide other criteria to distinguish the r.m.s. or m.a. of the residuals, etc., the only criterion to choose one pair from two or more is to calculate the mass of the system and compare it with that estimated from photometric and spectral data.

Period and semi-major axis, combined with parallax, yield the total mass of the system, according to Kepler's law:

$$M_d = M_1 + M_2 = a^3/(\pi^3 P^2) \qquad (1)$$

where P is the orbital period (in years), $M_{1,2}$ are the masses (in solar mass), a and π are the semi-major axis and the parallax (both in arcsec), respectively.

The masses of the components can also be estimated from observed photometry, trigonometric parallax, and a mass-luminosity relation. The appropriate formula is

$$M_{1,2} = f_{MLR} (m_{1,2} + 5 \lg \pi + 5 - A(l,b,\pi)), \qquad (2)$$

where $m_{1,2}$ are the apparent magnitudes, A is the interstellar extinction value, and $f_{MLR}$ is the mass-luminosity relation depending on the stellar luminosity class (LC). It is worth noting that the conformity of components to the mass-luminosity relation (in V band, in particular) is likely not exact, especially at the lower part of the main sequence, because of the stellar evolution and chemical abundance variations (Bonfils et al. 2005).

Another route requires spectral classification of both components and reliable mass-spectrum relation, $f_{MSR}$:

$$M_{1,2} = f_{MSR} (SpType_{1,2}). \qquad (3)$$

Let us call the masses, determined with Eqs. (1), (2) and (3), dynamical ($M_d$), photometric ($M_{ph}$), and spectral ($M_{sp}$), respectively. We computed the photometric mass for every pair and dynamical mass for every orbit. In the case of multiple orbits, we compared their values to choose the orbit that yields the most reliable mass.

For photometric mass estimation, we used the MLR of Malkov (2007) for upper-MS, Henry and McCarthy (1993) for lower-MS, and Henry et al. (1999) for lowest masses. Subgiants and early-type (O-F6) giants were considered to be 1 mag brighter than dwarfs (Halbwachs 1986). Lastly, for the few remaining late-giant and supergiant stars, photometric masses were estimated with Tables II and VI of Straizys and Kuriliene (1981). Pairs with unknown LC were considered to be MS-systems. The LC of secondary component, when unknown, was considered to be the same as for the primary. If the secondary magnitude was unknown (35 of 652 systems, see Appendix A), equal brightness of components was assumed, and, consequently, an upper limit for photometric mass was estimated. The photometric mass is absent in 17 cases, where the estimated absolute stellar magnitude is inappropriate for a given LC.

We also calculated the dynamical mass uncertainty. OARMAC does not contain uncertainties of P and a, and we estimated the following: $\sigma_P$ to be 5%, 10%, 20% for A, B, C quality grades, respectively, and $\sigma_a$ to be 3%, 6%, 12% for A, B, C quality grades, respectively.

As can be seen from Eq. (2), the absolute magnitude uncertainty depends mainly on the parallax uncertainty, and for the vast majority of our systems it does not exceed $0.^m3$. However, the main factor that contributes to the photometric mass uncertainty for a given absolute magnitude is the mass-luminosity relation accuracy. For intermediate mass MS-stars, the MLR slope and its standard deviation value (see, e.g., Malkov 2007) produce about 0.06 for the log M uncertainty (i.e. mass uncertainty is about 15%). For stars of other LCs, however, we estimate that uncertainty to be 2-3 times worse, mainly because of higher MLR standard deviation values.

Spectral masses were estimated from Table VI of Straizys and Kuriliene (1981). If the secondary spectral type was unknown, the listed spectral mass is the primary mass (i.e., it represents a minimum mass of the system). Main sequence is assumed if the luminosity class was unknown, which seems to be a reasonable assumption for our relatively nearby orbital binaries. Spectral mass is not available for six entries, which include Am, Be, L stars, and stars with inexact classification, e.g. M:. We estimated the spectral mass uncertainty to be about 20% for MS-stars (we refer to Straizys and Kuriliene (1981) figure 1) and about 30% for stars of other luminosity classes (see, e.g., Piskunov (1977) and the corresponding discussion in Straizys and Kuriliene (1981)).

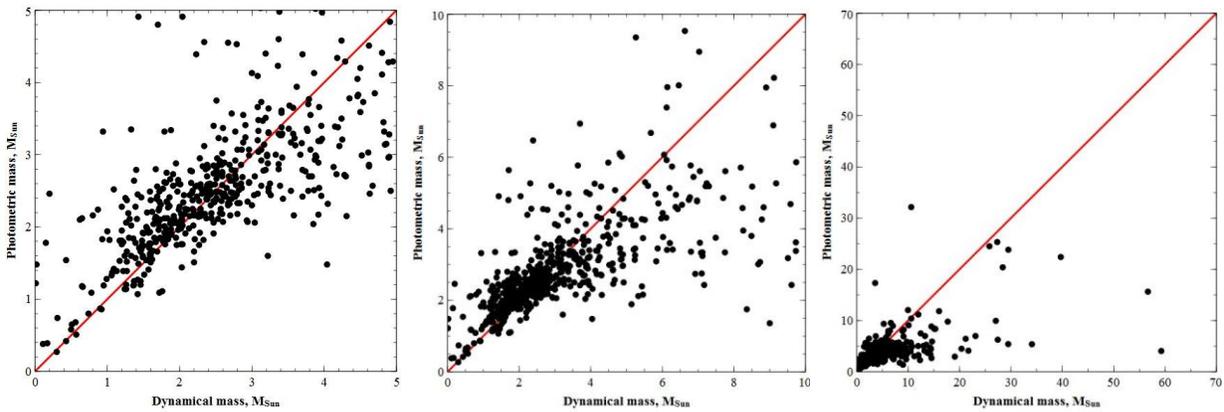

*Fig. 1 (a,b,c). Dynamical (Eq.1) versus photometric (Eq.2) mass of orbital binaries in different scales.*

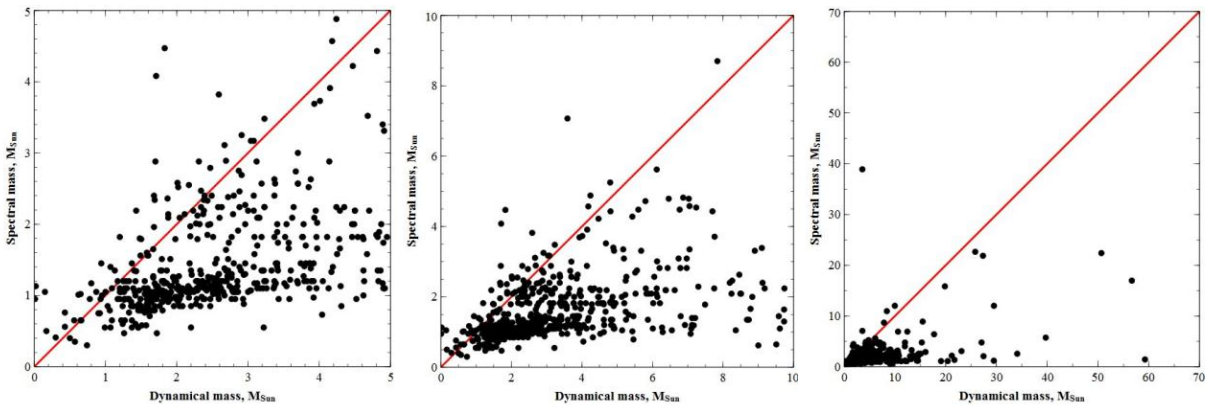

*Fig. 2 (a,b,c). Dynamical (Eq.1) versus spectral (Eq.3) mass of orbital binaries in different scales.*

In Figs 1-2 the stars we investigated have been plotted on the planes of the dynamical vs. photometric vs. spectral mass. First of all, we note that some systems exhibit obviously incorrect dynamical masses and/or a large discrepancy between dynamical, photometric and spectral masses (lists of such systems are given in Appendices C and D). The most obvious reason for that are poorly known parallaxes and/or orbital elements as well as the presence of a third still-undetected body (or subcomponents) that lead to dynamical mass overestimation (Tamazian et al. 2006). Among other reasons for a discrepancy between dynamical and photometric masses, we can mention incorrect spectral classification, interstellar extinction underestimation, and variability of components. These systems need further study. It should be noted that, for example, parallaxes of distant visual binaries can be refined by using relevant orbital and spectral data (Docobo et al. 2008). We should note also that Cvetkovic and Ninkovic (2010) in their study of Main Sequence binaries from ORB6 also found large deviations of the dynamical masses from the photometric ones for some systems.

All reasons for a discrepancy between masses, determined with Eqs. (1), (2) and (3), are listed in Table 1. An influence of input parameters to the determined masses is indicated. Observing a discrepancy between particular types of masses (dynamical, photometric or spectral) in Figs 1-2, one can estimate from Table 1, which input parameters have incorrect values.

| parameter | $M_d$ | $M_{ph}$ | $M_{sp}$ | parameter | $M_d$ | $M_{ph}$ | $M_{sp}$ |
|---|---|---|---|---|---|---|---|
| Spectral (temperature) class | no | no | yes | Semi-major axis, period | yes | no | no |
| Luminosity class | no | yes | yes | Variability | no | yes | yes/no[*] |
| Interstellar extinction | no | yes | no | Unresolved binarity | yes | yes | no |
| Parallax | yes | yes | no | Third body | yes | no | no |

*Table 1. Influence of various parameters and stellar characteristics on the values of dynamical, photometric and spectral masses. (\*) – depends on variability type.*

4. Statistical relations, parameter distributions and selection effects

The final list (so-called system list) contains 652 systems. This list is large enough for a reliable statistical analysis of this class of binary systems.

It is instructive to obtain relations between different observational parameters of orbital binaries. They can be used for statistical purposes. A period-axis relation is presented in Fig. 3.

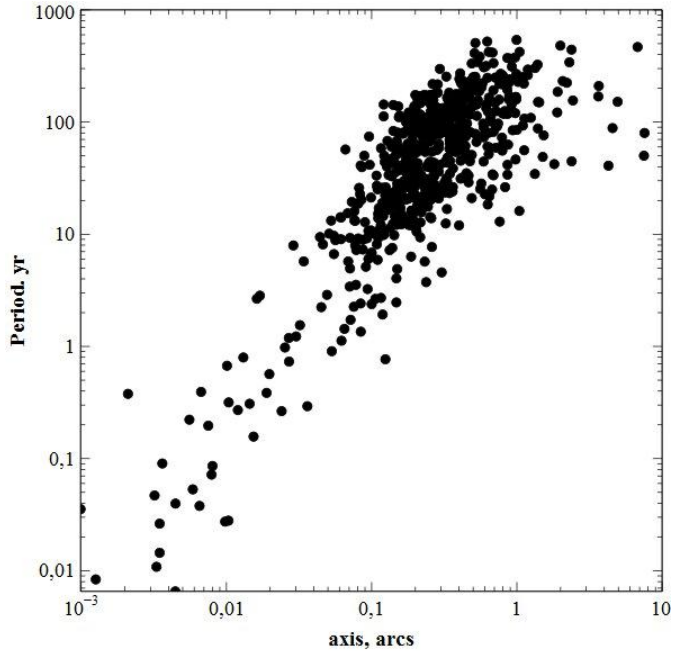

*Fig. 3. Statistical period – angular semi-major axis relation for orbital binaries.*

It can be seen from Fig. 3 that the majority of systems satisfy the following approximate relation: log P (y) = 1.35 log a" + 2, and it is valid for systems with a<1". These systems are relatively distant and, consequently, represent a mixture of masses, which broadens the relation. Systems with larger angular semi-major axes are nearby and the least massive systems, and both these circumstances lead to the flattening of the relation. In zeroth approximation one can consider all systems with a>1" to have periods of the order of 100 years.

It is well known that observational selection limits the possibilities of observation of binary systems of different types and can seriously distort the results of the analysis. Here we analyze the main selection effects that determine the sample of orbital binaries.

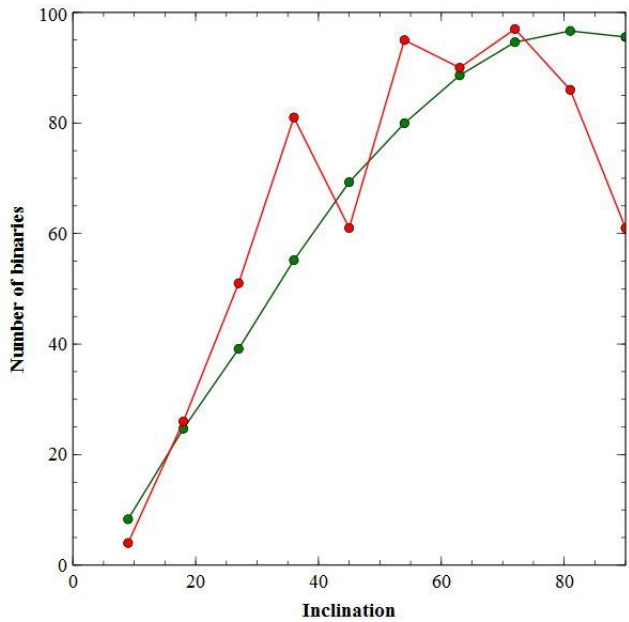

*Fig. 4. Distribution of stars with respect to the angle of orbital inclination I (red). Catalogued values in the 90-180 degrees range are converted in the 0-90 degrees range. The smooth green curve corresponds to the distribution in the case of random orientation of the planes of the orbits: dN ~ sin I di. Binary with an edge-on orbit is harder to discover, and it leads to a deficiency of highly inclined orbits (i around 90 degrees) in the figure.*

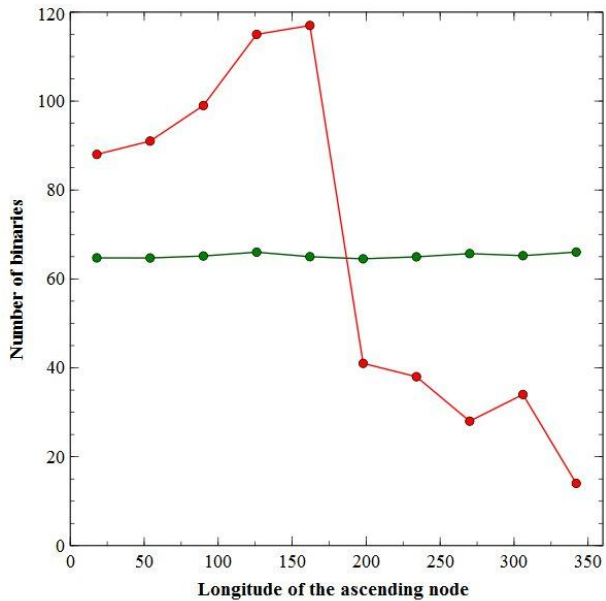

*Fig. 5. Distribution of stars with respect to the angle of longitude of the ascending node Ω (red). The horizontal green curve corresponds to the distribution in the case of random orientation of the planes of the orbits: dN ~ Ω dΩ. Number of orbits with Ω>180 in the figure is small due to the following reason. In the case of orbital binary without additional (e.g., radial velocity) data it is not possible to determine which node is ascending and which is descending, and in this case Ω is recorded in the 0-180 degrees (rather than 0-360 degrees) range. This tendency is apparently increasing toward Ω=180, as it can be seen in the figure (a slight increasing of the distribution between 0 and 180 degrees).*

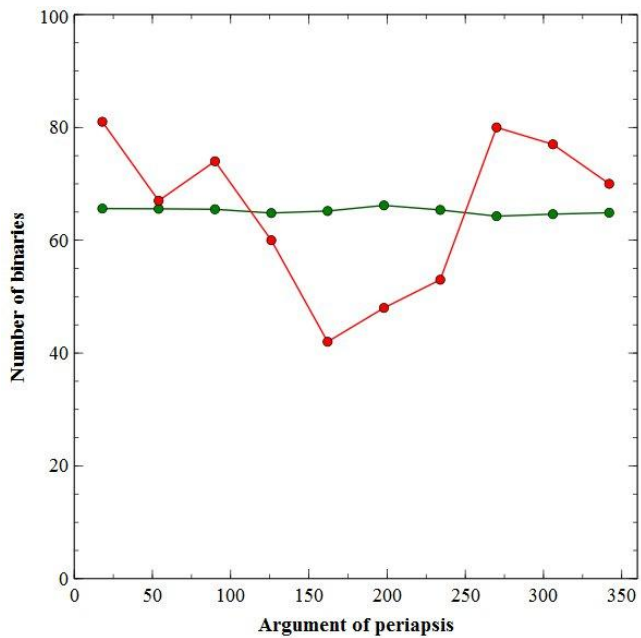

*Fig. 6. Distribution of stars with respect to the angle of argument of periapsis ω (red). The horizontal green curve corresponds to the distribution in the case of random orientation of the planes of the orbits: dN ~ ω dω. Maxima at ω around 90 and 270 degrees mean that binary observed from a "pointy" end is easier to discover than one observed from a "blunt" end (in the latter case a wide enough binary can be mistakenly considered as an optical pair and excluded from the statistics). It does not concern low-e or low-I orbits. Moreover for circular orbits (with eccentricity equal to zero) ω is undefined and, by convention, set equal to zero. Such orbits contribute to a peak around ω=0 in the figure.*

In Figs 4-6 we present distributions of 652 pairs from the system list along the Campbell elements: orbital inclination (i), longitude of the ascending node (Ω) and argument of periapsis (ω). It can be seen that i- and ω- distributions of catalogued orbital binaries are distorted by selection effects. Contrary, discoveries of orbits with different Ω are equiprobable events, however, Ω is more often recorded in the 0-180 degrees. Distributions of randomly aligned systems are given for comparison, they are modeled with the program described in (Malkov 2002).

The overwhelming majority of binaries in the system list have apparent magnitudes of the primary components m ≤ 9.$^m$5. Also, the majority of systems have semi-major axes a" ~ 0".1 -1": closer systems are not resolved by instruments with aperture D ~ 100 cm, which are generally used to detect visual binaries, and wider systems usually do not exhibit significant orbital motion.

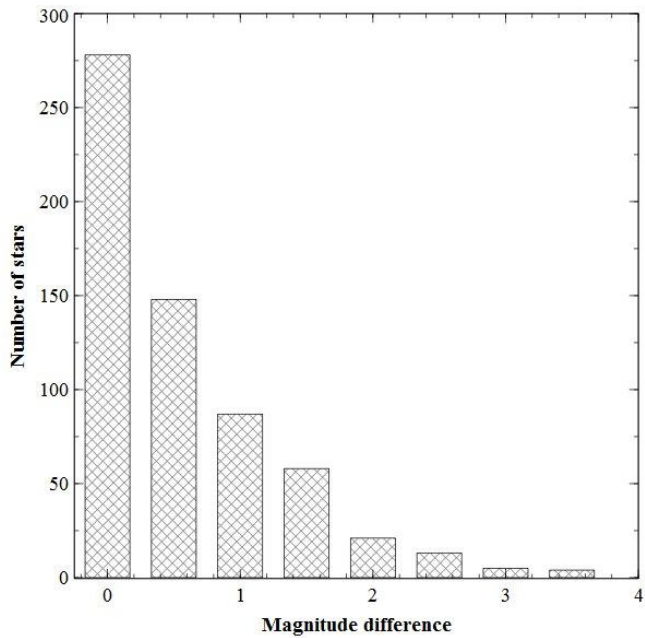

*Fig. 7. Distribution of stars with respect to the magnitude difference of the components dm. Few pairs with dm>4m and with unknown secondary brightness (see Appendix A) are not shown.*

The distribution with respect to the magnitude difference of the components dm (Fig. 7) gives, in principle, the possibility of estimating the mass ratio of the components. However, the dm distribution is also strongly distorted by selection. First, among the closest binaries, only pairs with comparable luminosities are detectable. Second, among binaries with the primary magnitude close to the limiting magnitude, only pairs with small dm are detected, and pairs with the secondaries fainter than the limiting magnitude appear as single stars and do not contribute to statistics.

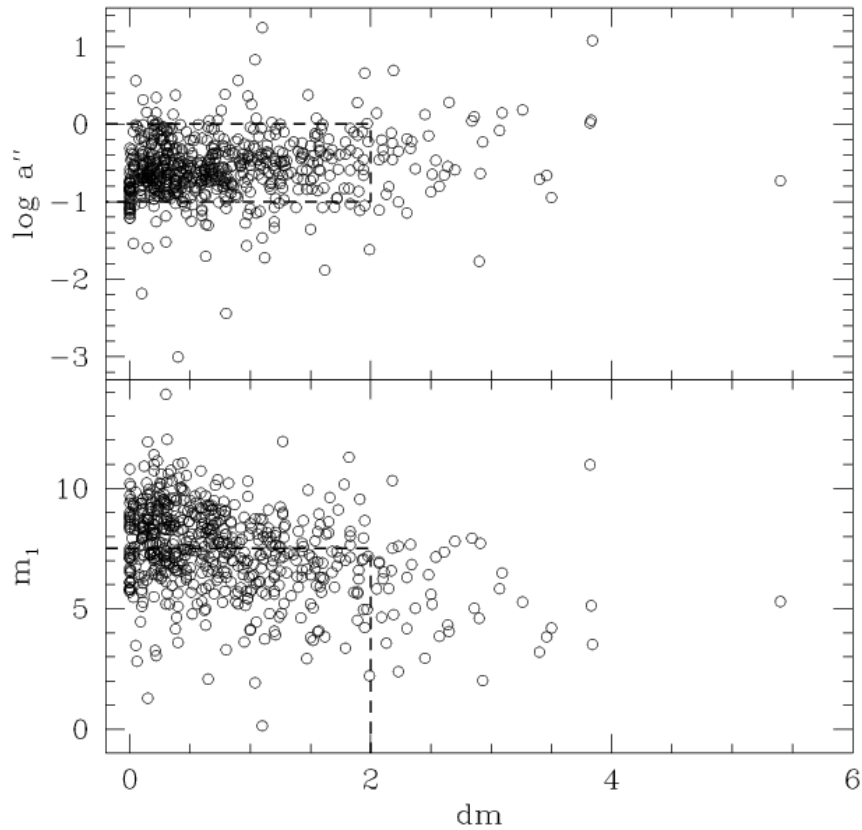

*Fig. 8. Difference between the magnitudes of the components vs the semi-major axis of the orbit (upper panel) and the magnitudes of the primaries (lower panel). The dashed lines mark the area that satisfies the definition of selected systems,*
*a" ~ 0".1 -1", $m_1 \leq 7.^m5$ and $dm \leq 2^m$.*

In Fig. 8, the stars we investigated have been plotted on the plane of the semi-major axis of the orbit (upper panel) and the magnitudes of the primaries (lower panel) vs the difference between the magnitudes of the components. It can be seen that the mean value of dm increases with increasing a" and increases with increasing brightness. One can see that, for systems with a" < 0".1, or a" > 1", or $m_1$ > $7.^m5$, or dm > $2^m$ the set is definitely incomplete. Note, however, that the restricted sample (systems with a" ~ 0".1 -1", $m_1 \leq 7.^m5$, dm ≤ $2^m$) surely does not include all visual binaries with corresponding restrictions posed on their magnitudes and angular separations. For example, a distant (d ~ 2.5 kpc) wide (a ~ 500 A.U) gravitationally bound system can be observed like a visual binary, but its orbital motion is so slow that it would not exhibit evident orbital motions and, hence, would not contribute to the sample as orbital binary.

For 207 systems of the 652 listed systems, a", m1 and dm satisfy these criteria. The distributions of these selected 207 systems, together with those of all 652 systems, among dynamical mass, period, semi-major axis, and eccentricity are shown in Figs. 9, 10, 11 and 12, respectively.

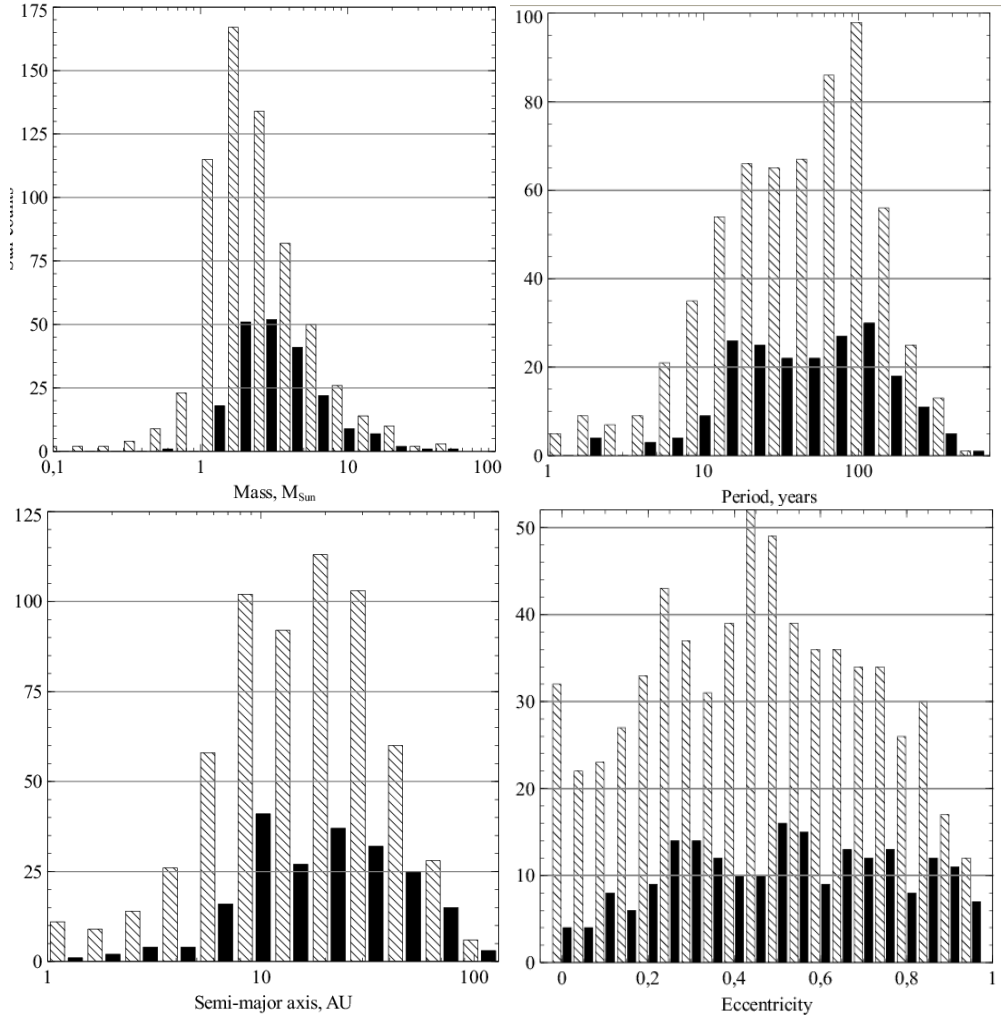

*Figs. 9-12. Dynamical mass, period, semi-major axis, and eccentricity distributions of all 652 listed systems (grey bars) and the selected 207 systems with*
*a" ~ 0".1 -1", $m_1 \leq 7.^m5$ and $dm \leq 2^m$ (black bars).*

The period distribution is shown in Fig. 10. Raghavan et al. (2010) surveyed the solar-type stars in the solar neighborhood and obtained log-normal distribution with a peak at about 300 years. Our sample is heavily biased for these long period binaries due to the lack of an extended set of observations.

Opik (1924) found that the semi-major axis distribution of binaries follows the f(a) ~ 1/a law. Poveda et al. (2007) confirmed this distribution for wide binaries (a>100 A.U.). Our sample is, inevitably, restricted at large a (see Fig. 11) due to long orbital periods of wide binaries, hence the number of binaries significantly drops at a>50 A.U. The distribution has a broad, nearly flat peak at 10<a<50 (which is consistent with the flat Opik distribution), and is steep again for close binaries with a<10 A.U. due to their small projected angular separations.

The eccentricity distribution (Fig. 12) is roughly flat for 0.3<e<1, but the small eccentricity pairs show evidence for Ambartsumian's (1937) f(e)=2e distribution.

It should be noted that the distributions (Figs. 9-12) are heavily distorted by the selection effects inherent to the OARMAC and ORB6 catalogues, and cannot be mistaken for initial or present-day distributions of binary parameters. They can be simulated using initial distributions on masses and orbital elements and involving stellar evolution and observational selection effects. However, before these effects are quantified, simulations cannot be compared to the catalogue and conclusions on the initial distributions cannot be made.

5. Conclusions

We collected and investigated visual binaries with known orbital parameters
to construct a list of orbital binaries from the OARMAC and ORB6 catalogues, with additional photometric and spectral information from WDS and SIMBAD. The resulting orbit list contains 3139 orbits of 2278 orbital binaries and represents the most comprehensive list to date of orbital binaries with known distances, photometry, spectral classification, and indication of the spectroscopic and eclipsing nature of the binary.

Next, we compiled a refined subset of the orbit list. The selection criteria used for the compilation of this system list were:
(1) the existence of trigonometric parallax,
(2) the absence of third companion, and
(3) the high quality of orbit.
For these systems, we selected and maintained in the system list only one (presumably, the best) orbit per system. The list contains 652 systems.

For these systems we calculated stellar masses independently from dynamical, photometric and spectral data. Comparison of these values shows (sometimes relatively large) discrepancy and can indicate a false input parameter. We also construct distributions of orbital binaries along observational parameters and discuss principal selection effects.

An analysis of selection effects shows that our subset can be considered to be complete in the following parameters area: semi-major axis a" ~ 0".1 -1", primary brightness $m_1 \leq 7.^m5$, and magnitude difference $dm \leq 2^m$. For 207 systems that satisfy these criteria we constructed distributions among dynamical mass, period, semi-major axis, and eccentricity. These distributions can be used to construct the initial mass function and star formation history of wide binaries. This is a subject for our future study. Our list of orbital binaries, containing orbital parameters from OARMAC and ORB6 as well as photometric and spectral information from WDS and SIMBAD, will be updated regularly, as all of the sources mentioned above are constantly updated. The list currently may be accessed from VizieR.
Some orbital binaries from our list need secondary magnitudes or radial velocity data. Lists of such binaries (see Appendices A and B, respectively) would be helpful for observers.

Acknowledgements

We thank Vakhtang Tamazian and Jose Docobo for their collaboration. This work was supported by the Russian Foundation for Fundamental Research grants 12-07-00528 and 12-02-31904, the Federal Science and Innovations Agency under contract 02.740.11.0247, the Presidium RAS program "Leading Scientific Schools Support" 3602.2012.2, and the Federal task program "Research and operations on priority directions of development of the science and technology complex of Russia for 2007-2012" (contract 16.518.11.7074).
This research has made use of the SIMBAD database, operated at the Centre de Donnees astronomiques de Strasbourg, the Washington Double Star Catalog maintained at the U.S. Naval Observatory, and NASA's Astrophysics Data System Bibliographic Services.

Appendices

A. Orbital binaries from the refined (system) list where secondary photometry is needed (format: WDS designation – visual magnitude of primary component)

| WDS name | V1 | WDS name | V1 | WDS name | V1 | WDS name | V1 |
|---|---|---|---|---|---|---|---|
| 00369+3343 | 4.36 | 05025+4105 | 3.75 | 16240+4822 | 10.44 | 20113-0049 | 3.23 |
| 00572+2325 | 4.42 | 05595+4457 | 1.90 | 16286+4153 | 2.79 | 20158+2749 | 4.52 |
| 01028+3148 | 5.51 | 07269+2015 | 5.90 | 17266-0505 | 4.54 | 20329+4154 | 7.09 |
| 01546+2049 | 2.64 | 09088+2638 | 5.98 | 17370+6845 | 4.80 | 20599+4016 | 6.56 |
| 02095+3459 | 3.00 | 09412+0954 | 3.52 | 18280+0612 | 5.73 | 21158+0515 | 3.92 |
| 02171+3413 | 4.85 | 11480+2013 | 4.53 | 18339+5144 | 8.07 | 22070+2521 | 3.81 |
| 02422+4012 | 4.91 | 12597-0349 | 5.80 | 18547+2239 | 4.59 | 22383+4511 | 6.40 |
| 02442-2530 | 7.00 | 14104+2506 | 4.82 | 19217-1557 | 4.58 | 22430+3013 | 3.00 |
| 04209+1352 | 6.16 | 16088+4456 | 4.23 | 19394+3009 | 4.70 | | |

B. We have compiled a list of orbital binaries that could use radial velocity measurements. Obviously, they should
- be massive (to cover intermediate and high mass regions, where the mass-luminosity relation based on empirical data of *eclipsing* binary components cannot be used to derive the stellar initial mass function, and, consequently, new observational data is needed, see Malkov (2003) for details),
- be close enough (to facilitate RV observations);
- have similar components (to be observed as an SB2 rather than an SB1 system).

Selected orbital binaries, needing RV data: dynamical mass is 3 solar mass or more, π>20 mas, dm<3$^m$ (format: WDS designation – dynamical mass – its error –visual magnitude of primary component – visual magnitude of secondary component – parallax – its error).

| WDS name | $M_d$ | $\sigma M_d$ | V1 | V2 | π | σπ |
|---|---|---|---|---|---|---|
| 00084+2905 | 5.19 | 0.16 | 2.22 | 4.21 | 33.62 | 0.35 |
| 02366+1227 | 19.08 | 0.98 | 5.68 | 5.78 | 28.79 | 0.43 |
| 04287+1552 | 4.46 | 0.28 | 3.74 | 4.86 | 21.69 | 0.46 |
| 06098-2246 | 3.13 | 0.32 | 6.56 | 6.57 | 20.39 | 0.69 |
| 07171-1202 | 4.05 | 1.71 | 8.61 | 6.96 | 21.60 | 2.11 |
| 07346+3153 | 5.43 | 0.96 | 1.93 | 2.97 | 64.12 | 3.75 |
| 09307-4028 | 3.70 | 0.75 | 3.91 | 5.12 | 53.15 | 0.37 |
| 10279+3642 | 3.37 | 1.05 | 4.62 | 6.04 | 21.19 | 0.50 |
| 11037+6145 | 5.56 | 0.33 | 2.02 | 4.95 | 26.54 | 0.48 |
| 11053-2718 | 3.79 | 0.14 | 5.70 | 5.70 | 23.13 | 0.29 |
| 11190+1416 | 3.04 | 1.02 | 7.01 | 7.99 | 32.53 | 1.39 |
| 12415-4858 | 7.30 | 3.02 | 2.82 | 2.88 | 25.06 | 0.28 |
| 13372-6142 | 3.39 | 1.41 | 5.98 | 7.22 | 27.99 | 0.58 |
| 15278+2906 | 3.08 | 0.24 | 3.68 | 5.20 | 29.17 | 0.76 |
| 15427+2618 | 4.18 | 0.28 | 4.04 | 5.60 | 22.33 | 0.50 |
| 17104-1544 | 7.05 | 0.49 | 3.05 | 3.27 | 36.91 | 0.80 |
| 18092-2211 | 4.04 | 1.32 | 9.58 | 9.92 | 30.49 | 1.20 |
| 19026-2953 | 5.24 | 0.37 | 3.27 | 3.48 | 36.98 | 0.87 |
| 21137+6424 | 3.85 | 0.23 | 7.21 | 7.33 | 23.39 | 0.42 |

C. Orbital binaries where difference between photometric and dynamical masses exceeds three-sigma limits (format: WDS designation – dynamical mass – its error – photometric mass).

| WDS name | $M_d$ | $\sigma M_d$ | $M_{ph}$ | WDS name | $M_d$ | $\sigma M_d$ | $M_{ph}$ |
| --- | --- | --- | --- | --- | --- | --- | --- |
| 00084+2905 | 5.19 | 0.18 | 4.10 | 11480+2013 | 3.97 | 0.25 | 4.97 |
| 00155-1608 | 0.17 | 0.02 | 0.39 | 13145-2417 | 0.94 | 0.48 | 3.32 |
| 00174+0853 | 0.02 | 0.02 | 1.48 | 13198+4747 | 1.76 | 0.13 | 1.11 |
| 00550+2338 | 1.86 | 0.14 | 2.29 | 14503+2355 | 0.11 | 0.01 | 0.38 |
| 00572+2325 | 3.88 | 0.22 | 5.02 | 15278+2906 | 3.08 | 0.24 | 4.09 |
| 01083+5455 | 0.01 | 0.01 | 1.22 | 16088+4456 | 3.38 | 0.29 | 4.98 |
| 02095+3459 | 4.11 | 0.18 | 5.26 | 16171+5516 | 1.72 | 0.13 | 1.09 |
| 02366+1227 | 19.08 | 2.57 | 2.95 | 16240+4822 | 0.50 | 0.02 | 0.58 |
| 02424+2001 | 3.64 | 0.47 | 5.77 | 17104-1544 | 7.05 | 0.51 | 3.61 |
| 03244-1539 | 0.62 | 0.11 | 2.10 | 17119-0151 | 0.31 | 0.14 | 0.74 |
| 03537+5316 | 1.00 | 0.24 | 1.94 | 17375+2419 | 13.76 | 1.34 | 3.71 |
| 04136+0743 | 2.17 | 0.14 | 2.77 | 17490+3704 | 1.84 | 0.14 | 2.27 |
| 05017+2050 | 0.15 | 0.04 | 1.78 | 18211+7244 | 1.71 | 0.06 | 2.07 |
| 05074+1839 | 0.65 | 0.09 | 2.12 | 19026-2953 | 5.24 | 0.37 | 3.45 |
| 05103-0736 | 1.20 | 0.18 | 2.28 | 19391+7625 | 1.24 | 0.07 | 1.60 |
| 05525-0217 | 1.29 | 0.28 | 2.19 | 19394+3009 | 7.49 | 0.77 | 3.75 |
| 06410+0954 | 3.58 | 1.59 | 17.32 | 19407-0037 | 1.88 | 0.45 | 3.34 |
| 07026+1558 | 0.87 | 0.33 | 2.24 | 19490+1909 | 2.23 | 0.35 | 4.39 |
| 07175-4659 | 0.43 | 0.23 | 1.54 | 19598-0957 | 1.54 | 0.13 | 2.00 |
| 07393+0514 | 2.03 | 0.05 | 1.44 | 20158+2749 | 2.39 | 0.73 | 6.47 |
| 07508+0317 | 1.70 | 0.37 | 4.80 | 20312+1116 | 0.20 | 0.38 | 2.46 |
| 09407-5759 | 5.83 | 0.53 | 3.09 | 21137+6424 | 3.85 | 0.55 | 2.04 |
| 10120-2836 | 1.43 | 0.86 | 4.91 | 22408-0333 | 0.80 | 0.22 | 2.16 |
| 11053-2718 | 3.79 | 0.19 | 2.70 | 23052-0742 | 4.62 | 0.55 | 2.86 |

D. Orbital binaries where difference between spectral and dynamical masses exceeds three-sigma limits (format: WDS designation – dynamical mass – its error – spectral mass).

| WDS name | $M_d$ | $\sigma M_d$ | $M_{sp}$ | WDS name | $M_d$ | $\sigma M_d$ | $M_{sp}$ |
| --- | --- | --- | --- | --- | --- | --- | --- |
| 00084+2905 | 5.19 | 0.18 | 3.09 | 10161-5954 | 3.77 | 0.39 | 1.82 |
| 00155-1608 | 0.17 | 0.02 | 0.50 | 11037+6145 | 5.56 | 0.80 | 2.75 |
| 00174+0853 | 0.02 | 0.02 | 1.13 | 11053-2718 | 3.79 | 0.19 | 1.41 |
| 00284-2020 | 2.14 | 0.32 | 1.05 | 11191+3811 | 5.89 | 0.40 | 2.09 |
| 00369+3343 | 11.98 | 2.06 | 4.79 | 11210-5429 | 15.29 | 3.02 | 4.79 |
| 00373-2446 | 1.79 | 0.29 | 0.91 | 11416+3145 | 2.56 | 0.30 | 1.25 |
| 00507+6415 | 9.12 | 1.74 | 2.40 | 11480+2013 | 3.97 | 0.25 | 1.66 |
| 00550+2338 | 1.86 | 0.14 | 1.35 | 12417-0127 | 2.69 | 0.06 | 2.89 |
| 00572+2325 | 3.88 | 0.22 | 2.63 | 13175-0041 | 3.39 | 0.31 | 1.45 |
| 01028+3148 | 8.19 | 0.87 | 2.40 | 13198+4747 | 1.76 | 0.13 | 0.85 |
| 01078-4129 | 3.16 | 0.34 | 2.09 | 13396+1045 | 3.41 | 0.36 | 1.40 |
| 01083+5455 | 0.01 | 0.01 | 0.95 | 13574-6229 | 2.48 | 0.38 | 1.12 |
| 01243-0655 | 2.59 | 0.41 | 1.30 | 14038-6022 | 29.50 | 5.38 | 12.02 |
| 01376-0924 | 2.51 | 0.43 | 1.13 | 14191-1322 | 4.50 | 0.22 | 2.19 |
| 01546+2049 | 3.19 | 0.13 | 1.82 | 14310-0548 | 1.89 | 0.22 | 0.95 |
| 02095+3459 | 4.11 | 0.18 | 1.82 | 14396-6050 | 1.98 | 0.03 | 1.79 |
| 02157+2503 | 3.09 | 0.30 | 1.17 | 14404+2159 | 6.55 | 1.65 | 1.05 |
| 02278+0426 | 1.37 | 0.12 | 0.65 | 14462-2111 | 2.51 | 0.21 | 1.02 |
| 02288+3215 | 1.46 | 0.20 | 0.55 | 15122-1948 | 6.20 | 0.56 | 2.82 |
| 02366+1227 | 19.08 | 2.57 | 1.13 | 16171+5516 | 1.72 | 0.13 | 0.47 |
| 02399+0009 | 2.83 | 0.56 | 1.05 | 16240+4822 | 0.50 | 0.02 | 0.40 |
| 02434-6643 | 14.55 | 4.04 | 1.20 | 16341+4226 | 6.05 | 1.14 | 2.57 |
| 02537+3820 | 3.88 | 0.60 | 1.38 | 16439+4329 | 1.27 | 0.16 | 0.65 |
| 03244-1539 | 0.62 | 0.11 | 1.01 | 17080+3556 | 3.09 | 0.14 | 1.82 |
| 03337+5752 | 2.47 | 0.38 | 1.13 | 17104-1544 | 7.05 | 0.51 | 4.58 |
| 03448+4602 | 1.54 | 0.13 | 0.79 | 17247+3802 | 2.37 | 0.30 | 0.95 |
| 03496+6318 | 3.76 | 0.61 | 1.82 | 17266-0505 | 2.59 | 0.21 | 1.30 |
| 04184+2135 | 2.96 | 0.47 | 1.45 | 17375+2419 | 13.76 | 1.34 | 2.09 |
| 04286+1558 | 4.14 | 0.35 | 2.88 | 17490+3704 | 1.84 | 0.14 | 1.05 |
| 04287+1552 | 4.46 | 0.29 | 2.00 | 17542+1108 | 4.83 | 1.03 | 1.20 |
| 04464+4221 | 2.53 | 0.35 | 1.05 | 18211+7244 | 1.71 | 0.06 | 1.13 |
| 05017+2050 | 0.15 | 0.04 | 1.05 | 18384-0312 | 2.42 | 0.32 | 1.07 |
| 05074+1839 | 0.65 | 0.09 | 1.02 | 18466+3821 | 1.47 | 0.12 | 0.88 |
| 05239-0052 | 2.74 | 0.51 | 1.13 | 19026-2953 | 5.24 | 0.37 | 2.46 |
| 05429-0648 | 2.77 | 0.28 | 1.30 | 19035-6845 | 3.21 | 0.54 | 1.10 |
| 06098-2246 | 3.13 | 0.32 | 1.17 | 19089+3404 | 1.51 | 0.20 | 0.58 |
| 06159+0110 | 4.62 | 0.81 | 1.35 | 19167-4553 | 1.24 | 0.20 | 0.60 |
| 06171+0957 | 3.13 | 0.24 | 2.09 | 19311+5835 | 1.59 | 0.07 | 0.79 |
| 06293-0248 | 0.30 | 0.02 | 0.41 | 19391+7625 | 1.24 | 0.07 | 0.85 |
| 06410+0954 | 3.58 | 1.59 | 38.90 | 19394+3009 | 7.49 | 0.77 | 1.78 |
| 06573-3530 | 2.53 | 0.16 | 1.19 | 19487+3519 | 2.53 | 0.32 | 1.17 |
| 07168+0059 | 2.08 | 0.35 | 0.85 | 19598-0957 | 1.54 | 0.13 | 1.10 |
| 07374-3458 | 6.74 | 1.03 | 3.09 | 20113-0049 | 10.09 | 0.66 | 2.88 |
| 07393+0514 | 2.03 | 0.05 | 1.27 | 20329+4154 | 1.58 | 0.05 | 0.88 |
| 07508+0317 | 1.70 | 0.37 | 2.88 | 20599+4016 | 2.18 | 0.15 | 1.10 |
| 07518-1354 | 1.90 | 0.09 | 1.02 | 21135+1559 | 3.96 | 0.66 | 1.82 |
| 07528-0526 | 2.20 | 0.23 | 1.35 | 21158+0515 | 4.68 | 0.20 | 3.52 |
| 07560+2342 | 1.82 | 0.22 | 1.05 | 21214+1020 | 2.60 | 0.44 | 1.13 |
| 07573+0108 | 2.85 | 0.37 | 1.13 | 21329+4959 | 6.12 | 1.01 | 2.24 |
| 08251-4910 | 1.60 | 0.14 | 0.85 | 21501+1717 | 2.56 | 0.14 | 1.51 |
| 08267+2432 | 3.54 | 0.80 | 1.13 | 21579-5500 | 5.24 | 0.83 | 1.58 |
| 08421-5245 | 1.98 | 0.24 | 0.98 | 22083+2409 | 2.58 | 0.33 | 1.00 |
| 08538-4731 | 4.25 | 0.50 | 1.58 | 22313-0633 | 2.02 | 0.13 | 1.13 |
| 09036+4709 | 6.63 | 1.41 | 2.19 | 22388+4419 | 2.26 | 0.16 | 1.07 |
| 09123+1500 | 1.80 | 0.25 | 0.91 | 22409+1433 | 2.17 | 0.33 | 0.97 |
| 09243-3926 | 4.29 | 0.76 | 1.66 | 23052-0742 | 4.62 | 0.55 | 1.79 |
| 09278-0604 | 2.06 | 0.31 | 1.00 | 23126+0241 | 3.43 | 0.64 | 0.95 |
| 09407-5759 | 5.83 | 0.53 | 2.09 | 23334+4251 | 1.84 | 0.20 | 1.05 |
| 09442-2746 | 6.27 | 1.23 | 1.13 | 23411+4613 | 2.16 | 0.28 | 1.13 |
| 09474+1134 | 3.57 | 0.49 | 1.66 | 23506-5142 | 2.28 | 0.37 | 0.95 |
| 10083+3136 | 2.77 | 0.22 | 1.25 | | | | |

Short description of the authors
Oleg Malkov works at the Institute of Astronomy as Head of Department of Physics of Stellar Systems. He is also a professor of astronomy at Moscow State University. Dmitry Chulkov has recently graduated from Faculty of Physics of Moscow State University and works at the Institute of Astronomy.